\documentclass[a4paper,11pt]{article}
\usepackage{pos}
\usepackage{subfig}
\usepackage{hyperref}
\hypersetup{urlcolor=black}
\title{Association between Recurrent Novae and \\ Nova Super-Remnants}
\ShortTitle{Association between Recurrent Novae and Nova Super-Remnants}

\author*{Michael William Healy-Kalesh}

\affiliation{Astrophysics Research Institute, Liverpool John Moores University, IC2 Liverpool Science Park,\\
146 Brownlow Hill, Liverpool, L3 5RF, United Kingdom}

\emailAdd{M.W.HealyKalesh@ljmu.ac.uk}

\abstract{Nova super-remnants (NSRs) are substantially extended structures (up to ${\sim}$130 parsecs across) encompassing recurrent novae. NSRs grow as a result of frequent nova eruptions transporting vast quantities of the locally surrounding interstellar medium away from the binary system over many millennia into a thin high-density shell, as the central white dwarf grows towards the Chandrasekhar limit. The prototypical NSR, first identified as such in 2014, is situated in the Andromeda Galaxy and belongs to the annually erupting nova, M\,31N 2008-12a (or `12a'). In this short review, modelling of evolving NSRs (including the 12a NSR) will be outlined as motivation towards searching for more of these phenomena in the Galaxy and beyond. The latest developments in this upcoming subfield of nova research will then be presented including the discovery of two new Galactic nova super-remnants (and their consequent modelling) and the first survey undertaken with the sole purpose of finding NSRs in the Andromeda Galaxy and the Large Magellanic Cloud.}

\FullConference{The Golden Age of Cataclysmic Variables and Related Objects - VI (GOLDEN2023)\\
 4-9 September 2023\\
Palermo - (Mondello), Italy\\}

\begin{document}
\maketitle

\section{Introduction}\label{sec:Introduction}
Novae are luminous eruptions arising from accreting white dwarfs (WD; \cite{1954PASP...66..230W,1995cvs..book.....W}). Within these binary systems, the WD accretes hydrogen-rich material either through Roche lobe overflow if a main sequence or subgiant donor, or via a high density wind if a red giant donor \cite{2012ApJ...746...61D}, leading to the formation of an accretion disk \cite{1976IAUS...73..155S,2016PASP..128e1001S}. As the accreting envelope grows, the temperature and pressure of the compressed matter at its base dramatically increases, such that the material reaches a state of electron degeneracy. Nuclear fusion ensues once the WD has accreted a critical amount of material. However, the degenerate nature of the envelope prohibits the matter from expanding in response to a significant rise in temperature: this leads to a thermonuclear runaway (TNR; \cite{1950AnAp...13..384S,1972ApJ...176..169S,1976IAUS...73..155S,1978A&A....62..339P,1981ApJ...243..926S}). The envelope heats during the TNR towards the Fermi temperature and becomes non-degenerate, resulting in the envelope expanding. Increasing opacity (as the material rapidly cools) and high expansion velocities (greater than the escape velocity of the WD) leads to the ejection of material into the surroundings and an expanding pseudo-photosphere -- this is the nova eruption \cite{1972ApJ...176..169S,1976IAUS...73..155S,1978A&A....62..339P}.

Observationally, this single nova eruption from a previously inconspicuous location would be defined a classical nova (CN). However, after the pseudo-photosphere has receded back to the surface of the WD, marking the end of the nova event and the transition to a period of quiescence \cite{1979A&A....72..192P}, accretion recommences onto the surviving \cite{2018ApJ...857...68H} (or re-establishing \cite{2007MNRAS.379.1557W}) accretion disk until the onset of a new nova eruption. The interval between successive eruptions defines the recurrence period ($P_{\text{rec}}$) of the system. While this process is predicted to occur in all novae \cite{1978ApJ...219..595F,2005ApJ...623..398Y,2016ApJ...819..168H,2018ApJ...860..110S}, only systems exhibiting more than one eruption are defined {\it observationally} as recurrent novae (RNe with $P_{\text{rec}} < 100$ years), though modifying the definition of such transients is warranted \cite{2022MNRAS.517.3864S}. Presently, thirty-four observationally-defined recurrent novae are recorded: ten in the Milky Way \cite{Darnley:2021ph} (though KT Eridani is a CN within the Galaxy exhibiting all the hallmarks of a RN (\cite{2014ApJ...788..164P,2022MNRAS.517.3864S,2024MNRAS.529..224S,2024MNRAS.529..236H} and see Section~\ref{sec:KT Eridani Nova Super-Remnant}), four in the Large Magellanic Cloud (LMC) and twenty in the Andromeda Galaxy (M31) \cite{2020AdSpR..66.1147D,2024MNRAS.528.3531H}. Out of this collection of RNe, systems displaying the shortest recurrence periods possess a massive WD  being fuelled by an elevated accretion rate ($\dot{M}$, \cite{1988ApJ...325L..35S}).

As first proposed by Whelan \& Iben \cite{1973ApJ...186.1007W}, growing a carbon-oxygen (CO) WD to the Chandrasekhar limit ($\text{M}_{\text{Ch}}$; \cite{1931ApJ....74...81C}) within a binary -- akin to a nova system -- can produce a type Ia supernova (SN Ia). Specifically for the nova scenario, material within the envelope not expelled during the eruption will be retained and thereby increase the mass of the WD; this mechanism is reinforced by a large collection of theoretical studies (see e.g. \cite{2005ApJ...623..398Y,2007ApJ...659L.153H,2016ApJ...819..168H}). Novae, therefore, take their place among the collection of prospective SN Ia progenitors as one of the feasible single degenerate pathways and are the brightest of all proposed progenitors, even during their quiescent state \cite{Darnley:2021ph}.

Material expelled during a nova eruption (up to $10^{-4} \ \text{M}_{\odot}$) at velocities ranging from hundreds to thousands of kilometers per second \cite{2001IAUS..205..260O} expands outwards to form a nova shell \cite{1983ApJ...268..689C,1985ApJ...292...90C,1990LNP...369..179W,1995MNRAS.276..353S}. Remnants forming around novae with high-velocity ejecta (`fast novae') may adopt a spherical structure with evident fragmentation whereas `slow novae' may produce ellipsoidal shells with equatorial and tropical rings alongside polar blobs of emission \cite{1995MNRAS.276..353S}. Moreover, the equatorial rings and asphericity of such shells can be employed to determine the orbital inclination of the binary \cite{1995MNRAS.276..353S,2020MNRAS.499.2959H} and rotation of the WD \cite{1998MNRAS.296..943P}, respectively. Nova shells from singular nova eruptions have been identified around many CNe including DQ\,Herculis \cite{1978ApJ...224..171W,2007MNRAS.380..175V}, T\,Aurigae \cite{1980ApJ...237...55G,1995MNRAS.276..353S}, HR\,Delphini \cite{2003MNRAS.344.1219H}, DO\,Aquilae \cite{2020MNRAS.499.2959H} to name a few, with {\it sub-parsec sizes} being a key commonality amongst these structures. Nova shells have also been discovered around three dwarf novae (DNe), namely Z Camelopardalis \cite{2007Natur.446..159S,2012ApJ...756..107S} (recently a second larger shell has been detected around this DN by \cite{2024MNRAS.529..212S}), AT Cancri \cite{2012ApJ...758..121S,2017MNRAS.465..739S} and a DN associated with the nebula Te 11 \cite{2016MNRAS.456..633M}, thereby proving that old novae can indeed transform into DNe, as predicted by the hibernation scenario of cataclysmic variables \cite{1986ApJ...311..163S}. Additionally, a nova shell encircling the Galactic {\it recurrent} nova T Pyxidis reveals evidence for interacting ejecta from separate eruptions \cite{1997AJ....114..258S,2013ApJ...768...48T,2015ApJ...805..148S,2023arXiv231204277I}.

This article is based upon an invited review talk entitled `{\it Association between Recurrent Novae and Nova Super-Remnants}' presented at The Golden Age of Cataclysmic Variables and Related Objects VI in Palermo, Italy (September 2023) and is structured in the following way. Section~\ref{sec:The prototypical Nova Super-Remnant} introduces the discovery and initial modelling of the prototypical nova super-remnant surrounding the RN, M\,31N 2008-12a, from which all the following work was built upon. Section~\ref{sec:Predicting a Likely Abundance of Nova Super-Remnants} details an extensive suite of simulations of growing nova super-remnants employed to determine the impact various system parameters have on their long-term structure. Section~\ref{sec:KT Eridani Nova Super-Remnant} and Section~\ref{sec:RS Ophiuchi Nova Super-Remnant} presents two new recently discovered Galactic nova super-remnants, and Section~\ref{sec:An Apparent Dearth of Nova Super-Remnants in M31 and the LMC} focusses on the first survey of M\,31 and the LMC specifically searching for more nova super-remnants.

\section{The Prototypical Nova Super-Remnant}\label{sec:The prototypical Nova Super-Remnant}
Residing in the Andromeda Galaxy, the most rapidly recurring nova known is M\,31N 2008-12a (referred to as `12a'; see \cite{2020AdSpR..66.1147D,2016ApJ...833..149D,2017ASPC..509..515D,2018ApJ...857...68H}). Since its discovery in 2008 \cite{2008Nis}, this remarkable RN has been seen in eruption annually -- with a mean $P_{\text{rec}} = 0.996 \pm 0.030$ yrs \cite{2020AdSpR..66.1147D,2024MNRAS.528.3531H} --- and is the only nova for which future eruptions can be predicted. Such frequent eruptions occur as a result of the near-Chandrasekhar mass WD (${\simeq} \ 1.38\,\text{M}_{\odot}$;  \cite{2015ApJ...808...52K}) accreting material from its red giant donor via Roche lobe overflow \cite{2017ApJ...849...96D} at a substantial rate of $(0.6 \lesssim \dot{M} \lesssim 1.4) \times 10^{-6} \ \text{M}_{\odot} \ \text{yr}^{-1}$ \cite{2017ApJ...849...96D}. Furthermore, its likely CO WD \cite{2017ApJ...849...96D,2017ApJ...847...35D} is growing towards the $\text{M}_{\text{Ch}}$ and is predicted to detonate as a SN Ia in less than 20,000 years \cite{2017ApJ...849...96D}.

\begin{figure}
\centering
\includegraphics[width=\columnwidth]{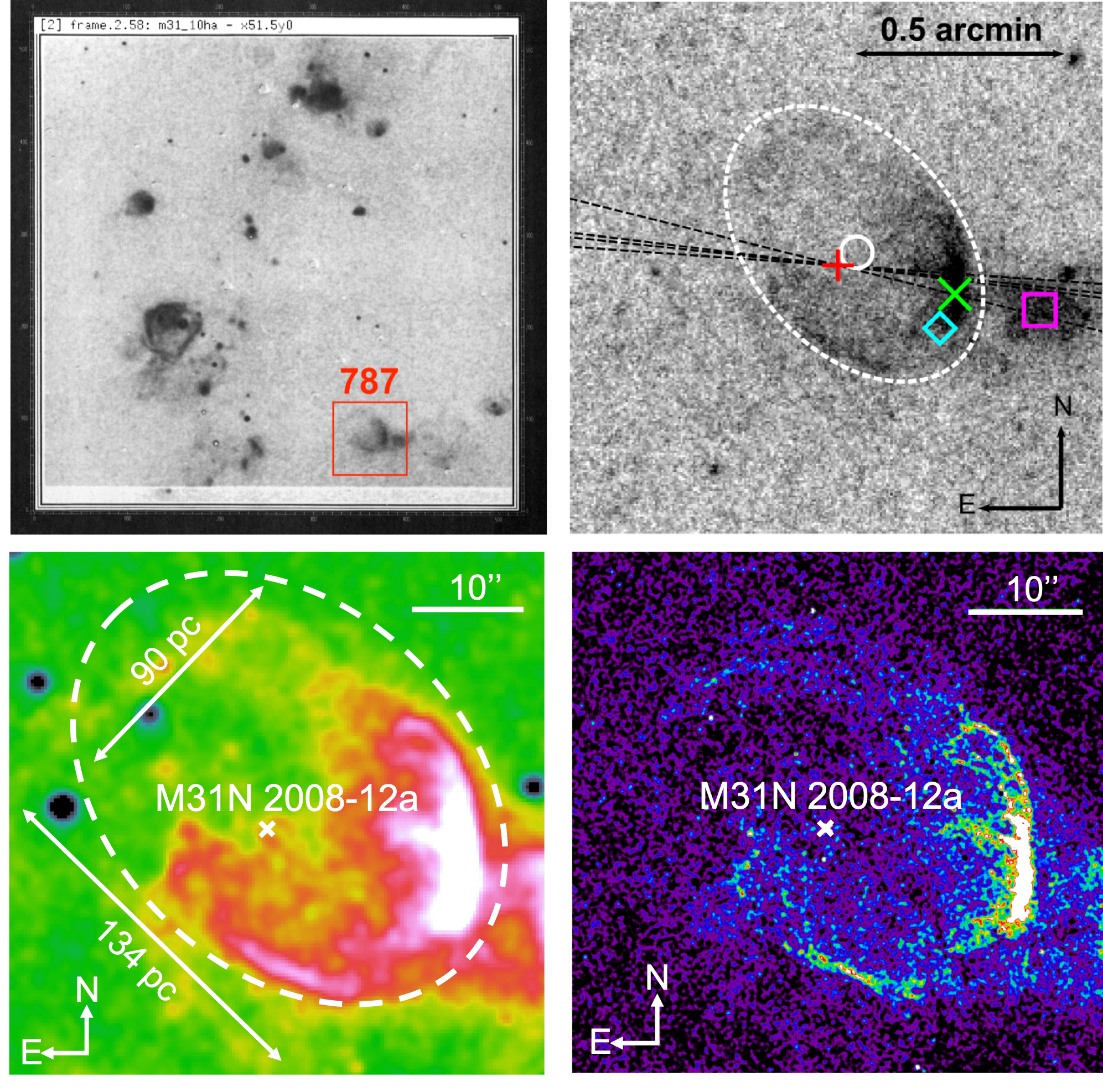}
\caption{{\it Top left panel:} Narrow-band H$\alpha$ image of the region of M\,31 containing the NSR around 12a (designated object 787 in this work). Taken and adapted from Walterbos \& Braun \cite{1992A&AS...92..625W}, reproduced with permission \copyright \ ESO. {\it Top right panel:} Continuum-subtracted H$\alpha$ image of the NSR surrounding 12a taken with the Liverpool Telescope. The dashed white line approximately follows the border of the elliptical remnant. Taken from Darnley et al. \cite{2015A&A...580A..45D}, reproduced with permission \copyright \ ESO. {\it Bottom left panel:} Liverpool Telescope continuum-subtracted H$\alpha + $[N II] image of the 12a NSR with its dimensions provided. Based on figure from Darnley et al. \cite{2019Natur.565..460D} and recreated with the same data. {\it Bottom right panel:} Continuum-subtracted {\it Hubble Space Telescope} H$\alpha + $[N II] image of the NSR surrounding 12a. The high-resolution of {\it HST} now reveals the bright emission from the fragmented material on the western side of the NSR. Based on figure from Darnley et al. \cite{2019Natur.565..460D} and recreated with the same data. \label{fig:12a NSR all}}
\end{figure}

However, the extreme nature of this rapidly recurring nova is not confined to its configuration and annual eruptions. First associated with 12a in Darnley et al. \cite{2015A&A...580A..45D}, this RN was found to be encompassed by a vastly extended (unprecedented in size) elliptical remnant (see top right panel of Figure~\ref{fig:12a NSR all}) spanning 90 by 134 parsecs \cite{2015A&A...580A..45D}, orders of magnitude larger than other nova shell. Previous narrow-band surveys covering this area of M\,31 (\cite{1992A&AS...92..625W,2008ApJ...686.1261C,2012ApJ...760...13F} -- see the top left panel of Figure~\ref{fig:12a NSR all}) had already uncovered this nebulosity but it was not attributed to the (at the time) unknown central nova. It was only once the recurring nature of 12a became apparent that archival data was searched (to find any previously missed eruptions) leading to the `rediscovery' of the structure. Follow-up observations of the remnant with the Liverpool Telescope \cite{2004SPIE.5489..679S} (see bottom left panel of Figure~\ref{fig:12a NSR all}) and a search through Local Group Galaxy Survey (LGGS; \cite{2007AJ....133.2393M}) data confirmed this finding. High-spatial resolution {\it Hubble Space Telescope} H$\alpha$ + [NII] imaging was able to unveil the finer structure of the shell and deep spectroscopy with the Gran Telescopio Canarias and Hobby-Eberly Telescope ruled out the possibility of the structure being a supernova remnant \cite{2019Natur.565..460D} -- heralding the discovery of {\it first ever nova super-remnant} \cite{2019Natur.565..460D}.

Darnley et al. \cite{2019Natur.565..460D} employed one-dimensional hydrodynamic simulations (using the \texttt{Morpheus} code \cite{2007ApJ...665..654V}) to determine if sustained recurrent nova eruptions from 12a could create the unprecedented size of the observed NSR. Modelling 100,000 annual identical 12a-like eruptions sweeping out the pre-populated interstellar medium (ISM) indeed led to the formation of a vast excavated region surrounding the central nova consisting of three distinct regions: a low-density inner cavity, an ejecta pile-up (a region continually shock-heated by the ejecta from new eruptions) and a high-density shell \cite{2019Natur.565..460D}. At the end of the full simulation, 100,000 annual eruptions had swept up 17\,$\text{M}_{\odot}$ of ISM (3,000$\times$ the mass ejected by the nova), thereby demonstrating that NSRs are predominantly made up of ISM material, with any nova-like material being almost completely diluted \cite{2019Natur.565..460D}. Extrapolating the outer and inner edge of the simulated NSR shell to the dimensions of the observed NSR (maintaining a shell thickness of $\sim$22\%, found to be consistent with the observations) illustrated that it would require annual eruptions for $6 \times 10^{6}$ years to reach its current size and would sweep up a total mass of $3 \times 10^{4} \ \text{M}_{\odot}$ in the process \cite{2019Natur.565..460D}.

On account of the most massive WD known in any nova system \cite{2015ApJ...808...52K} and ${\sim}63\%$ mass accumulation efficiency \cite{2015ApJ...808...52K}, 12a is {\it the leading} single degenerate progenitor of a SN Ia. Combining these properties and status with the vast NSR hosted by 12a, created through frequent nova eruptions occurring for millions of years, indicates that NSRs could be persistent features surrounding other RNe with near-$\text{M}_\text{{Ch}}$ WDs primed to explode as a SN Ia \cite{2019Natur.565..460D,2020AdSpR..66.1147D}. Furthermore, the formation of the NSR entails the removal of almost exclusively H-rich ISM away from the central WD prior to explosion, thereby providing a reason for the lack of hydrogen in SN Ia spectra \cite{1997ARA&A..35..309F,2016sros.confE.137H,Darnley:2021ph}.

\section{Predicting a Population of Nova Super-Remnants}\label{sec:Predicting a Likely Abundance of Nova Super-Remnants}

The NSR encompassing 12a exists as a direct result of the massive WD and significant accretion rate powering frequent highly-energetic nova eruptions. Therefore, Darnley and Henze \cite{2020AdSpR..66.1147D} postulated that similar NSRs should accompany other RNe as these systems will share similar attributes to 12a (high mass WD, high $\dot{M}$, short $P_{\text{rec}}$). Identifying these other extended NSRs could reveal locations of previously unknown, unconfirmed or `extinct' RNe (systems in which the donor has lost all donable material) and signify the single degenerate progenitor of past or upcoming SN Ia \cite{Darnley:2021ph}. But how can the characteristics (e.g. accretion rate and WD mass) of RN systems, as well as the local environment, impact the growth and long-term evolution, and therefore observability, of their NSR?

Building on previous modelling \cite{2019Natur.565..460D}, an extensive suite of 1D hydrodynamic simulations  \cite{2023MNRAS.521.3004H} following evolving eruptions from RNe going on to form NSRs provided key insights into answering this question. The system parameters and ranges implemented within these simulations, guided by nova outburst models from Yaron et al. \cite{2005ApJ...623..398Y}, consisted of accretion rate ($10^{-9} - 10^{-7} \ \text{M}_{\odot} \ \text{yr}^{-1}$); ISM density ($0.1 - 100$ H cm$^{-3}$); initial WD mass ($0.65 - 1.3 \ \text{M}_{\odot}$) and WD temperature ($10^{7} - 3 \times 10^{7}$ K). Across the majority of the runs, the WD was grown from its initial mass up to the Chandrasekhar limit (a temporal upper limit was set for systems with low accretion rates as the WD eroded) with the evolving properties of individual eruptions and the (shortening) recurrence period taken as the input for the modelling \cite{2023MNRAS.521.3004H}. This allowed structures to grow around different RNe thereby providing a representative sample of the potential wider NSR population.

\begin{figure}
\centering
\includegraphics[width=\columnwidth]{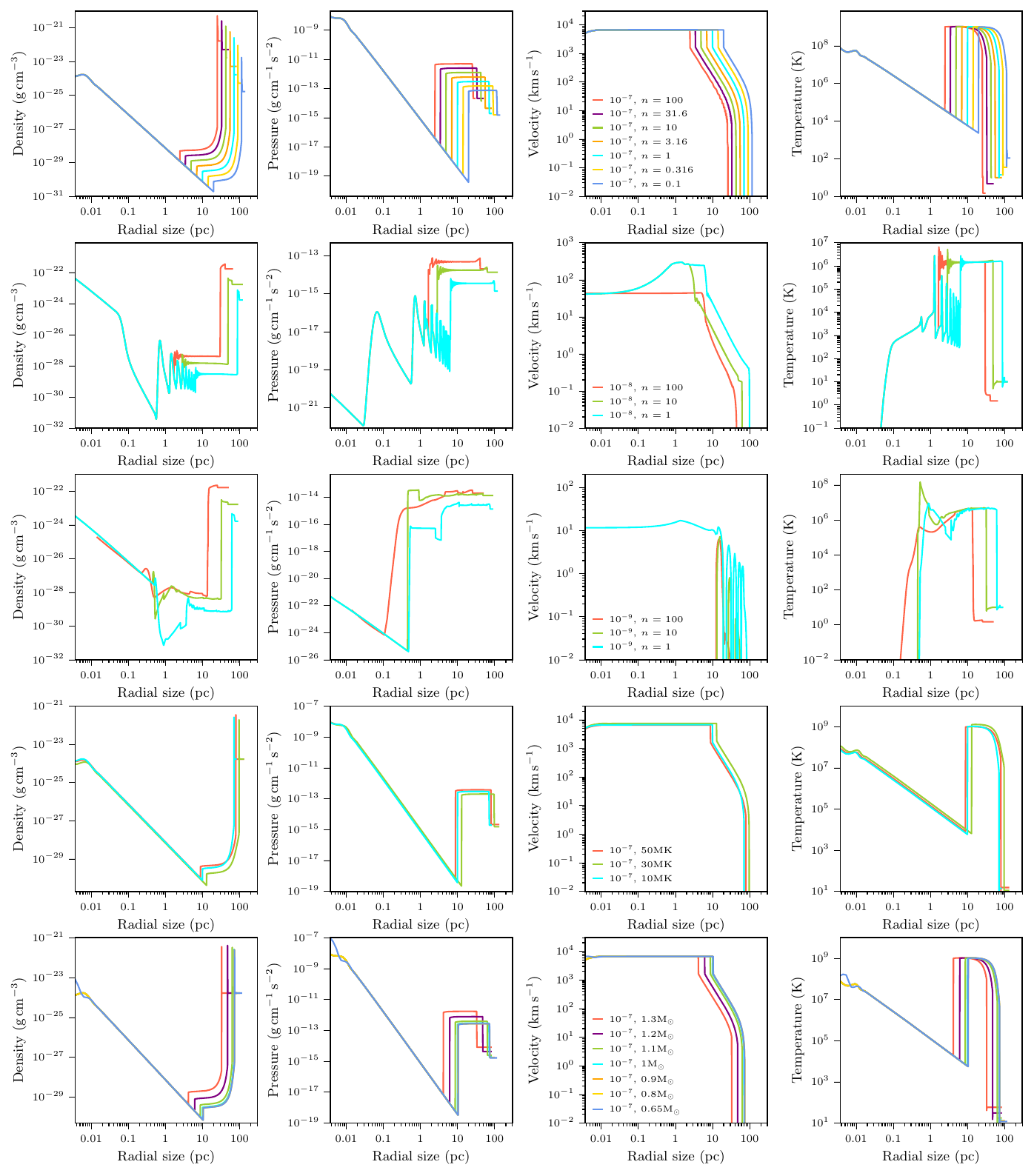}
\caption{Dynamics of the final epoch of grown NSRs from Healy-Kalesh et al. \cite{2023MNRAS.521.3004H}. {\it First row}: NSRs grown with accretion rate of $10^{-7} \, \text{M}_{\odot} \, \text{yr}^{-1}$ with fixed WD temperature and initial mass for various ISM densities. {\it Second row}: Same as first row but with accretion rate of $10^{-8} \, \text{M}_{\odot} \, \text{yr}^{-1}$. {\it Third row}: Same as first row but with accretion rate of $10^{-9} \, \text{M}_{\odot} \, \text{yr}^{-1}$. {\it Fourth row}: NSRs grown with fixed accretion rate, ISM density and initial WD mass for various WD temperatures. {\it Fifth row}: NSRs grown with fixed accretion rate, ISM density and WD temperature for various WD initial masses. Figure 7 from Healy-Kalesh et al. \cite{2023MNRAS.521.3004H}. \label{fig:all sims}}
\end{figure}

Regardless of the combination of parameters used, all modelled RNe grew a NSR (as shown in Figure~\ref{fig:all sims}), including novae with long inter-eruption periods and even those systems with an eroding WD (opening up the possibility of NSRs existing around old novae with low mass WDs). To reiterate this key finding: {\it all novae should be surrounded by a NSR} \cite{2023MNRAS.521.3004H}. Though, only NSRs hosted by novae with high accretion rates will be currently observable. Akin to Darnley et al. \cite{2019Natur.565..460D}, Healy-Kalesh et al. \cite{2023MNRAS.521.3004H} found that nova systems with high mass accretion rates ($10^{-7} \ \text{M}_{\odot} \ \text{yr}^{-1}$) can create NSRs (up to ${\sim}120$ parsecs in radius) consisting of a very low density cavity surrounded by a hot ejecta pile-up region, all contained within a thin, high density shell \cite{2023MNRAS.521.3004H}. Though, less energetic ejecta combined with longer intervals early in the evolution of the Healy-Kalesh et al. \cite{2023MNRAS.521.3004H} runs (allowing for prolonged phases of radiative cooling and energy loss) significantly altered the structure of the remnant shells compared to the NSR shell in Darnley et al. \cite{2019Natur.565..460D}, such that they were substantially thinner and denser. For the range of parameters considered, higher ISM densities and higher $\dot{M}$ both restricted the radial size of the growing NSR whereas the temperature and initial mass of the WD both had a relatively small influence on the evolution of the NSR \cite{2023MNRAS.521.3004H}. Notably, NSRs grown from oxygen-neon (ONe) WDs (initial WD mass $> 1.1 \ \text{M}_{\odot}$) were found to be considerably reduced in size (see bottom row of Figure~\ref{fig:all sims}), making these types of NSRs more difficult to find.

A simulated NSR grown from a nova with similar attributes as 12a (namely $\dot{M}$ and ISM density) extended to a comparable radial size (${\sim}71$\,pc) as the observed NSR (67\,pc) and adopted a reminiscent structure: a low density cavity (inferred in the observed NSR through a lack of emission from its central region \cite{2019Natur.565..460D}); a very hot ejecta pile-up region (potentially evidenced by the inner `knot' close to 12a and the strong [O III] emission there indicating high temperatures \cite{2019Natur.565..460D}); a cool, dense shell (a lack of [O III] emission there signifies the shell has cooled below the ionisation temperature of O$^{+}$) and an approximate structure for the H$\alpha$ emission \cite{2023MNRAS.521.3004H}. The majority of this H$\alpha$ emission, as with the majority of X-ray flux, originates from the inner edge of the NSR shell where newly-impacting ejecta leads to elevated levels of collisional excitation.

The study of Healy-Kalesh et al. \cite{2023MNRAS.521.3004H} concludes by stating `{\it detectable remnants can form around novae with very different system parameters and so should actively be searched for around all types of novae, not just those with very short recurrence periods}'. With ${\sim}$350 Galactic novae (\cite{2017ApJ...834..196S} and see \href{https://asd.gsfc.nasa.gov/Koji.Mukai/novae/novae.html}{`Koji's List of Recent Galactic Novae'}), over 1100 nova candidates in M\,31 (\cite{2010AN....331..187P} and see \href{https://www.mpe.mpg.de/~m31novae/opt/m31}{`Optical Novae and candidates in M 31'}), and the considerable number of such novae predicted to be recurrent \cite{2014ApJ...788..164P,Darnley:2021ph}, there is a substantial list of targets to search around for more nova super-remnants. Although, the substantial challenge facing this hunt stems from their very low surface brightness, and for Galactic examples, their vast projected size on the sky.

\section{KT Eridani Nova Super-Remnant}\label{sec:KT Eridani Nova Super-Remnant}

The Condor Array Telescope (Condor; \cite{2023PASP..135a5002L}) is a new facility with a large enough ($2.3 \times 1.5$ deg$^2$) field-of-view to capture a Galactic NSR in its totality and sensitive enough (with long exposures permitted through a dedicated observing campaign) to reach down to surface brightnesses NSRs are predicted to exhibit -- in this respect, Condor is a true ``nova super-remnant detector''.

Motivated by the existence of the NSR surrounding 12a, Shara et al. \cite{2024MNRAS.529..224S} opted to point Condor towards the long suspected eleventh Galactic RN, KT Eridani. An observing campaign from October 2021 to November 2022 uncovered a shell-like structure surrounding KT Eri, bright in H$\alpha$ and extending ${\sim}0.6$ deg$^2$ which, if at the distance of KT Eri, corresponds to a projected diameter of ${\sim}$50 pc. On account of its substantial size compared to other nova shells (see Section~\ref{sec:Introduction}) and its reminiscence to the 12a NSR, along with the elimination of a supernova remnant origin, Shara et al. \cite{2024MNRAS.529..224S} declared the discovery of the {\it second nova super-remnant}. Furthermore, the existence of a NSR associated with KT Eri reinforced the claim that it is the eleventh member of the Galactic RN collection \cite{2024MNRAS.529..224S} and strengthened the utility of NSRs for certifying unconfirmed RNe \cite{Darnley:2021ph}.

Hydrodynamic modelling demonstrated that a NSR of this size could be grown around KT Eri in ${\sim}$50,000 years \cite{2024MNRAS.529..236H}, which is in agreement with the nova still being located within its own shell as it will have traversed ${\sim}$0.29 shell diameters in this time period \cite{2024MNRAS.529..224S}. The modelled NSR had a shell thickness of ${\sim}14\%$ shell which matches the northern edge boundary of the observed structure \cite{2024MNRAS.529..224S,2024MNRAS.529..236H}, though, owing to the expected complexity of the surrounding ISM, the KT Eri NSR does not have a well-defined shell. Moreover, the NSR is seemingly compressed in the direction of KT Eri's proper motion, possibly indicating interaction of the outer NSR boundary with a previously created bow shock \cite{2024MNRAS.529..236H}.

\section{RS Ophiuchi Nova Super-Remnant}\label{sec:RS Ophiuchi Nova Super-Remnant}
With RS Ophiuchi being a Galactic RN, and arguably the most well-studied nova after 12a, it represents another worthy candidate for hosting a NSR. Modelling from Healy-Kalesh et al. {\cite{2023MNRAS.521.3004H}} demonstrated that, for an estimated ISM density, the potential NSR shell around RS Oph would have dimensions of $\gtrsim100$\,pc with a cavity $\gtrsim16$\,pc, equating to $\gtrsim250^{\prime}$ and $\gtrsim40^{\prime}$, respectively \cite{2024MNRAS.529L.175H}. As the cavity exists due to an almost complete removal of all cool gas from the inner regions of the NSR, a search through far-infrared IRIS (Improved Reprocessing of the {\it IRAS} Survey) data \cite{2005ApJS..157..302M} proved to be fruitful. A cavity-like feature was promptly found with dimensions (if co-located with RS Oph) of ${\sim}16 \times 5$ pc, matching the predicted cavity size \cite{2023MNRAS.521.3004H}, and so was claimed to be likely associated with the central RN \cite{2024MNRAS.529L.175H}; the same feature was independently identified over a decade before \cite{2008ASPC..401...90V}, though the wind from RS Oph was suggested to be responsible for its creation.

Instead, the claim of a NSR cavity \cite{2024MNRAS.529L.175H} was given further credence through the independent discovery of a H$\alpha$-bright NSR shell. With the success of unveiling the second NSR around KT Eri, Condor was used to target RS Oph. Observing the surroundings of RS Oph with lengthy exposures revealed another vast structure, again bright in H$\alpha$ and [N II] emission. At a distance of RS Oph, the uncovered nebulosity measures ${\sim}$30 pc across and so takes its place as the {\it third nova super-remnant} discovered (Michael Shara, presented during this conference and private communication). Not only did this discovery of a H$\alpha$-bright NSR shell around RS Oph reinforce the connection between RNe and NSRs but it also complimented and ultimately strengthened the separate discovery of the NSR-like feature found within archival far-IR imaging presented in Healy-Kalesh et al. \cite{2024MNRAS.529L.175H}.

\section{An Apparent Dearth of Nova Super-Remnants in M31 and the LMC}\label{sec:An Apparent Dearth of Nova Super-Remnants in M31 and the LMC}
With two new nova super-remnants being unveiled surrounding the Galactic RNe, KT Eridani and RS Ophiuchi, the hunt for further examples of these structures is assuredly underway. While Galactic searches with Condor and other upcoming facilities will doubtlessly increase the sample size of NSRs, attention must also be turned towards our Local Group neighbours. With a combined collection of 24 RNe, M\,31 (20) and the LMC (4) are ideal surveying sites for more NSRs \cite{2024MNRAS.528.3531H}. As with the first identified NSR around 12a, any super-remnant in M\,31 or the LMC hosted by a RN there would span similar dimensions (approximately tens of arcseconds) on the sky.

A search through narrow-band (H$\alpha$ and [S II]) LGGS \cite{2006AJ....131.2478M,2007AJ....134.2474M} data expectedly recovered the NSR hosted by 12a but did not reveal structures surrounding the other RN in M\,31. Likewise, Faulkes Telescope South narrow-band (H$\alpha$ and [S II]) imaging of the environs of the four RNe in the LMC also did not display anything akin to a super-remnant \cite{2024MNRAS.528.3531H}. While a simple explanation for this {\it apparent} dearth of NSRs in M\,31 and the LMC could be that these other RNe do not own such a structure, predictions of their existence from modelling and the unequivocal existence of the three NSRs already uncovered would suggest otherwise. Instead, the lack of NSRs could result from many of the systems being considered containing an ONe WD; such novae would have short $\text{P}_{\text{rec}}$ from birth and so create a relatively small super-remnant in the time taken to reach $\text{M}_{\text{Ch}}$ \cite{2023MNRAS.521.3004H}. Alternatively, other NSRs will be intrinsically fainter than the shell owned by 12a on account of longer $\text{P}_{\text{rec}}$ ($> 1$ year) and potentially sparser surroundings \cite{2024MNRAS.528.3531H}, hampering detection.

Crucially, NSR luminosity estimates from modelling \cite{2023MNRAS.521.3004H} predict that the second brightest nova super-remnant hosted by a known RN (M\,31N 2017-01e \cite{2017ATel10042....1W,2022RNAAS...6..241S}) would be ${\sim}1.25$ magnitudes fainter than 12a's NSR. In fact, Healy-Kalesh et al. \cite{2024MNRAS.528.3531H} found that the {\it NSR hosted by 12a is only just on the cusp of visibility} and would not be detectable in LGGS data if it were approximately one magnitude fainter. This is the moral of the story for current NSR research -- much deeper exposures of the greatly-extended environments of novae are imperative for uncovering more examples of NSRs. The strategy moving forward is therefore clear, albeit logistically challenging.

\section{Conclusion}
Moving forward, the strategy for finding more NSRs is four-fold: (i) continue the search around systems in M\,31 and the LMC  (see Section~\ref{sec:An Apparent Dearth of Nova Super-Remnants in M31 and the LMC}) with deeper observations and within different wavelength regimes (H$\alpha$, X-ray and infrared); (ii) increase the number of novae to search around by including RNe `masquerading' as classical novae (see, for example, \cite{2014ApJ...788..164P}); (iii) conduct surveys of other Local Group galaxies that can fit within one field-of-view of ground-based telescopes where there exists a population of novae to catch any clearly associated structures; and (iv) observe the surroundings of other RNe inevitably uncovered by upcoming next generation wide-field surveys such as Vera C. Rubin Observatory \cite{2019ApJ...873..111I}. 

To date, modelling has predominantly been employed to study the evolution and structure of growing nova super-remnants \cite{2019Natur.565..460D,2023MNRAS.521.3004H,2024MNRAS.529..236H}; this is research that would benefit from simulations with higher dimensionality, for example, to investigate complex ISM interactions. Yet, at the latter stages of NSR growth, a more extraordinary event is possible if the central WD were to explode as a SN Ia. As briefly explored in \cite{HealyPhD}, simulations of this violent epoch can provide insights into the observational signatures of such an event to be searched for -- a finding that would firmly cement RNe as SN Ia progenitors.

Recently Darnley \& Henze \cite{2020AdSpR..66.1147D} queried ``{\it whether M\,31N 2008-12a and its super-remnant is a rare, or even unique, phenomenon; or is it just the tip of the iceberg?}'' While far from settled, the answer to this question is now becoming clearer with the recent discoveries of NSRs associated with the Galactic RNe, KT Eridani and RS Ophiuchi, to place alongside the prototype surrounding the extragalactic RN, M\,31N 2008-12a. The search for more nova super-remnants is very much underway and thus the answer to this question will become better quantified as the sample of NSRs grows.

\acknowledgments
\noindent Firstly, I would like to thank the Scientific and Local Organising Committees of The Golden Age of Cataclysmic Variables and Related Objects VI, in particular Franco Giovannelli, Rosa Poggiani and Francesco Reale, for inviting me to attend their conference to talk about nova super-remnants and for their hospitality during the whole event. I would like to thank Professor Matt Darnley for reading over this manuscript and providing very valuable comments and suggestions. I also wish to thank the anonymous referee for their feedback. Finally, I want to acknowledge the funding by the UK Science and Technology Facilities Council (STFC) for my current PDRA position and my prior PhD studentship, in which time all of my research on nova super-remnants has taken place.

\bibliographystyle{JHEP}
\bibliography{bibliography}

\newpage
\noindent {\bf DISCUSSION}

\bigskip
   
\noindent {\bf (Q) Mumbua (Miriam) Nyamai:} Why is H$\alpha$ emission preferred when observing nova super-remnants in recurrent novae?
\vspace{2mm}

\noindent {\bf (A) Michael Healy-Kalesh:} Observing nova super-remnants in H$\alpha$ is preferred because the whole structure, and in particular the high-density shell, is predominantly made up of swept-up interstellar medium (mostly consisting of hydrogen). Continual ionisation of this H-rich material by energetic nova ejecta and the consequent recombination contributes to their H$\alpha$ luminosities.

\bigskip
\bigskip
\noindent {\bf (Q) Ken Shen:} What is more important for the formation of the super-remnant? The shorter recurrence time of energy and mass injection, or the higher effective rate of energy and mass injection? In other words, would you get the same super-remnant if all the mass were ejected at once?
\vspace{2mm}

\noindent {\bf (A) Michael Healy-Kalesh:} The large scale structure of the grown nova super-remnant is dependent on the total kinetic energy injected into the surrounding interstellar medium from the central nova. But the way in which the ejecta is structured largely impacts the formation of the nova super-remnant. One single bulk ejecta would sweep out the surrounding ISM and retain its thickness, likely resulting in a NSR structured with a very-low density cavity contained within a thick high-density shell, devoid of a pile-up region. Whereas, in the work outlined in this article, the total mass is ejected in discrete eruptions. Radiative cooling ensures that the thinner shell compresses and loses its thickness throughout the evolution and a pile-up region can form behind the expanding shell. Additionally, the observability of the NSR would be dramatically impacted if all of the mass were ejected at once. The low-mass ejecta from frequent eruptions impacting the forming NSR ionises the material which then recombines such that the remnant is relatively bright. Whereas, one bulk ejecta would sweep up the ISM and would not be ionised again from subsequent eruptions.

\end{document}